\documentclass[aps,prl,reprint, superscriptaddress,showpacs]{revtex4-1}

\usepackage{dcolumn}
\usepackage{bm}

\usepackage{graphicx}
\usepackage{url}

\date{Received 9 February 2017; published 11 May 2017}

\begin{document}

\preprint{APS/123-QED}


\title{Quantum Fluctuations along Symmetry Crossover\\
in a Kondo-correlated Quantum Dot} 

\author{Meydi Ferrier}
\email{meydi.ferrier@u-psud.fr}
\affiliation{Department of Physics, Graduate School of Science, Osaka University, 560-0043 Osaka, Japan.}
\affiliation{Laboratoire de Physique des Solides, CNRS, Univ. Paris-sud, Universit{\'e} Paris Saclay, 91405 Orsay Cedex, France.}

\author{Tomonori Arakawa}
\affiliation{Department of Physics, Graduate School of Science, Osaka University, 560-0043 Osaka, Japan.}

\author{Tokuro Hata}
\affiliation{Department of Physics, Graduate School of Science, Osaka University, 560-0043 Osaka, Japan.}

\author{Ryo Fujiwara}
\affiliation{Department of Physics, Graduate School of Science, Osaka University, 560-0043 Osaka, Japan.}

\author{Rapha{\"e}lle Delagrange}
\affiliation{Laboratoire de Physique des Solides, CNRS, Univ. Paris-sud, Universit{\'e} Paris Saclay, 91405 Orsay Cedex, France.}

\author{Richard Deblock}
\affiliation{Laboratoire de Physique des Solides, CNRS, Univ. Paris-sud, Universit{\'e} Paris Saclay, 91405 Orsay Cedex, France.}

\author{Yoshimichi Teratani}
\affiliation{Department of Physics, Osaka City University, Osaka 558-8585, Japan.}

\author{Rui Sakano}
\affiliation{The Institute for Solid State Physics, The University of Tokyo, Chiba 277-8581, Japan.}

\author{Akira Oguri}
\affiliation{Department of Physics, Osaka City University, Osaka 558-8585, Japan.}

\author{Kensuke Kobayashi}
\email{kensuke@phys.sci.osaka-u.ac.jp}
\affiliation{Department of Physics, Graduate School of Science, Osaka University, 560-0043 Osaka, Japan.}
\affiliation{Center for Spintronics Research Network (CSRN), Graduate School of Engineering Science, Osaka University, Osaka 560-8531, Japan.}

\begin{abstract}
Universal properties of entangled many-body states are controlled by their symmetry and quantum fluctuations. By the magnetic-field tuning of the spin-orbital degeneracy in a Kondo-correlated quantum dot, we have modified quantum fluctuations to directly measure their influence on the many-body properties along the crossover from $SU(4)$ to $SU(2)$ symmetry of the ground state. High-sensitive current noise measurements combined with the nonequilibrium  Fermi liquid theory clarify that the Kondo resonance and electron correlations are enhanced as the fluctuations, measured by the Wilson ratio, increase along the symmetry crossover.  Our achievement demonstrates that nonlinear noise constitutes a measure of quantum fluctuations that can be used to tackle quantum phase transitions.
\end{abstract}

\maketitle

{\it Introduction.}---Understanding the emergence of universal properties in entangled many-body states is a major task in various branches of physics. The key challenge is to unveil how they are governed by quantum fluctuations. The Kondo effect~\cite{Kondo1964,HewsonBook} is one of the paradigms for such many-body states, arising from entanglement of a localized electron with conduction electrons that screen its magnetic moment. It plays an important role in transport through quantum dots~\cite{Goldhaber-Gordon1998,Cronenwett1998,Schmid1998}, where the dot and conducting electrons are entangled in a singlet ground state with $SU(2)$ symmetry to screen the localized spin $S=\frac{1}{2}$. Interestingly, when several degrees of freedom including orbital magnetic moment as well as spins are combined in a highly degenerate internal moment, more peculiar Kondo many-body states are formed~\cite{Coqblin1969,LeHur2007} with different symmetries~\cite{Nozieres1980} because of the resulting rich spin-orbital configurations. At the heart of these phenomena are the quantum fluctuations between different configurations reflecting quantum uncertainty. However, the evolution of fluctuations between different symmetries of the Kondo states remains unexplored.

In this Letter, by tuning the Kondo state in a carbon nanotube (CNT) with a magnetic field~\cite{NygardNature2000}, we  continuously change the quantum fluctuations to directly measure their influence on the many-body properties. Nonequilibrium current noise measurements along the crossover between $SU(4)$  and $SU(2)$ symmetry of the ground state quantitatively demonstrate how fluctuations affect the residual interaction between quasiparticles to enhance the Kondo resonance~\cite{Mora2009,Sakano2011,Sakano2006,Filippone2014a}. 
This work demonstrates an unambiguous link between the effective charge $e^*$ and quantum fluctuations, and suggests that it can be extended in regions of broken symmetries where no theory exists yet. Hence it provides a new way to measure quantum fluctuations via the effective charge $e^*$ in the nonlinear noise~\cite{Gogolin2006,Sela2006}, which can be used to unveil their critical role in quantum phase transitions. 

{\it Principle of the symmetry crossover.}---In CNT quantum dots, electrons possess spin and orbital (valley) degrees of freedom as shown in the central inset of Fig.~1. Each state is fourfold degenerate yielding the $SU(4)$ Kondo ground state~\cite{SasakiPRL2004,ChoiPRL2005,PabloNature2005,DelattreNatPhys2009,Laird2014a}. We induce a never-observed crossover from the $SU(4)$ to the $SU(2)$ symmetry at half filling (two electrons in the dot) by tuning the orbital and spin degeneracy with a magnetic field. As the degeneracy of the screened degree of freedom decreases, quantum fluctuations between two of its components induce a larger relative change in the total magnetic moment. This enhancement of the fluctuations, which is eventually quantified  by the Wilson ratio, leads to a stronger residual interaction between quasiparticles~\cite{Mora2009,Sakano2011a}. We investigate this scenario by probing the conductance, Kondo temperature, and nonequilibrium effective charge $e^*$, which are precisely compared with the Fermi liquid theory extended out of equilibrium~\cite{Oguri2001,Glazman2005condmat,Oguri2005a}. The originality of this work lies in the control of the many-body symmetry at constant filling of the dot, which ensures that fluctuations are the only variable and enables us to demonstrate the continuous crossover between the two symmetries.

\begin{figure}[htbp]
\center
\includegraphics[width=\linewidth]{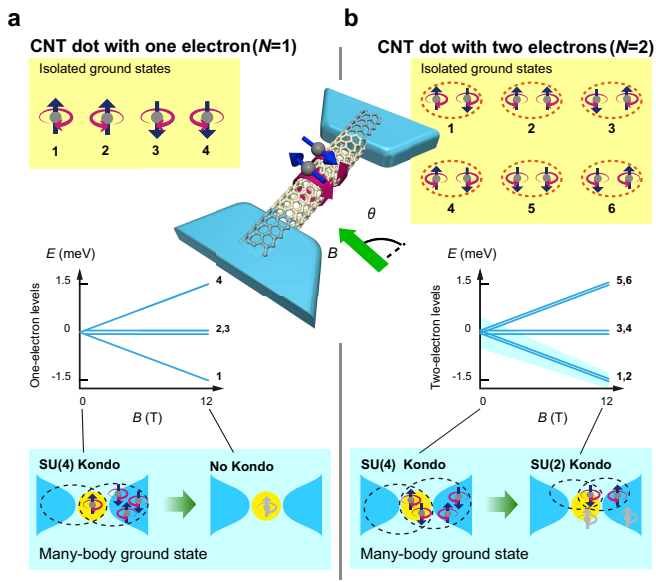}
\caption{Central inset: the spin and orbital (valley) degrees of freedom of an electron in a CNT are shown by the straight and circle arrows, respectively. The magnetic field is applied making an angle $\theta$ with the CNT axis. The figure is partially drawn by VESTA~\cite{MommaJAC2011}. (a) Top panel: representation of the four degenerate ground states for a dot containing one particle ($N=1$ and $N=3$). Middle panel: single-particle energy spectrum as a function of $B$ for the case of $g_{\rm orb}\mu_B\cos\theta=g_S\mu_BS$. Bottom panel: at $B=0$, four states are degenerate forming an $SU(4)$ Kondo state. At finite field, the Kondo effect disappears. (b) Top panel: representation of the six degenerate ground states for a dot with two particles ($N=2$).  Middle panel: the total magnetic energy is $E=(\sigma_1+\tau_1+\sigma_2+\tau_2)\mu_BB$. Each line is twice degenerate and corresponds to the states labeled on the graph. The shade around the ground state represents $k_BT_K$. Bottom panel: at $B=0$, an $SU(4)$ Kondo state is formed due to the sixfold degeneracy. At high field it evolves continuously to $SU(2)$ (see text).}
\end{figure}

More specifically, consider a CNT in the magnetic field making an angle $\theta$ with its axis (see the central inset of Fig.~1). The field $B$ acts on the spin with the Zeeman energy $E^{\rm spin}=\frac12\sigma g_S\mu_BB$, whereas only its parallel component $B\cos\theta$ acts on the orbital momentum~\cite{Laird2014a} adding a term $E^{\rm orb}=\tau g_{\rm orb}\mu_BB\cos\theta$. Here, $\mu_B$ is the Bohr magneton, $g_S$ and $g_{\rm orb}$ are the spin and orbital Land\'e factors,  $\sigma=\pm1$ refers to the spin direction, and $\tau=\pm1$ refers to the valley quantum number. Our experiment was carried out with $g_{\rm orb}\approx 4$ and $\theta\approx 75^\circ$~\cite{SupMat}, which satisfies a special condition of  $g_{\rm orb}\cos\theta\approx g_S/2=1$.

The top panel of Fig.~1(a) shows the four degenerate ground states for a dot containing one particle ($N=1$ corresponds to  a single electron case and $N=3$ to a single hole case). When the magnetic field is applied, energy shift is almost the same ($\sim\mu_BB$) for both the spin and valley because  $g_{\rm orb}\cos\theta\approx g_S/2=1$, and the total magnetic energy in this case is $E=(\sigma + \tau)\mu_BB$ [see the middle panel of Fig.~1(a)]. Hence, four states are degenerate forming an $SU(4)$ Kondo state at $B=0$, while the degeneracy is lifted and the Kondo effect disappears for finite $B$ [bottom panel of Fig.~1(a)]. 

When the dot contains two electrons ($N=2$), there are six degenerate states at $B=0$ as shown in the top panel of Fig.~1(b). Because $g_{\rm orb}\cos\theta\approx g_S/2=1$ again, as $B$ increases, two states are shifted by $\Delta E=-2\mu_BB$, two are unaffected, and the last two are shifted by $\Delta E=+2\mu_BB$ [see the middle panel of Fig.~1(b)]. As a result, the ground state remains doubly degenerate for finite $B$. At $B=0$ the six degenerate states give rise to the $SU(4)$ Kondo effect. When the field is sufficiently high that the ground state is only doubly degenerate within the characteristic energy scale $k_BT_K$ [shaded area of the middle panel of Fig.~1(b)], the $SU(2)$ Kondo state emerges. A partial representation of the many-body ground state that screens the magnetic moment on the dot is given in the bottom panel of Fig.~1(b). The full ground state is given in Ref.~\cite{SupMat}. This crossover for the $N=2$ case is addressed in this study.

{\it Experimental setup.}---Our device~\cite{Ferrier2015a} is a CVD grown CNT~\cite{Kasumov2007} on a nondoped silicon wafer connected to metallic pads composed of a bilayer Pd($6\ $nm)/Al($70\ $nm). A side gate electrode can tune the discrete energy levels of the CNT quantum dot. The conductance and the noise measurements were performed for the device placed in the dilution fridge, whose base temperature was 16 mK.  A small in-plane magnetic field of $0.08$~T in addition to $B$ was always applied to suppress the superconductivity of Al. For the noise measurement, the device was connected to an $LC$ circuit with a resonance frequency of 2.58~MHz thermalized on the mixing chamber of the fridge~\cite{Arakawa2013}. The power spectral density of the noise was obtained by amplifying the noise signal with a homemade cryogenic amplifier fixed on the 1K pot, taking the time-domain signal by a digitizer, and performing the fast Fourier transformation of the data. The current noise of the dot was extracted from the fit of the shape of the resonance curve in the frequency domain. The analysis for the measured noise was done in the same way as in Ref.~\cite{Ferrier2015a}. 

{\it Observation of $SU(4)$-$SU(2)$ Kondo crossover.}---The stability diagram of our CNT quantum dot for $B=0$, 4, 8, and 10~T is shown in Fig.~2(a) as a contour plot of the differential conductance $G$ as a function of the source-drain voltage ($V_{\rm sd}$) and gate voltage ($V_g$). The dot filling consists of successive shells containing four almost degenerate states and we label by $N=0,1,2,$ and $3$ the number of electrons in the last occupied shell. $N$ is electrostatically controlled by $V_g$. The Kondo resonance manifests itself as a maximum in $G$ at $V_{\rm sd}=0$, which appears as a bright vertical line parallel to the $V_g$ axis, called the Kondo ridge. At $B=0$~T, the resonance is seen for every filling as expected for the $SU(4)$ Kondo effect~\cite{Makarovski2007,Ferrier2015a}. For $N=1$ and $3$, the zero bias conductances are  $G\approx G_Q$, while $G=1.85 G_Q$ for  $N=2$, where $G_Q \equiv \frac{2e^2}{h}$ ($e$ is the electronic charge  and $h$ is the Planck constant) is the conductance quantum.

\begin{figure}
\center
\includegraphics[width=1\linewidth]{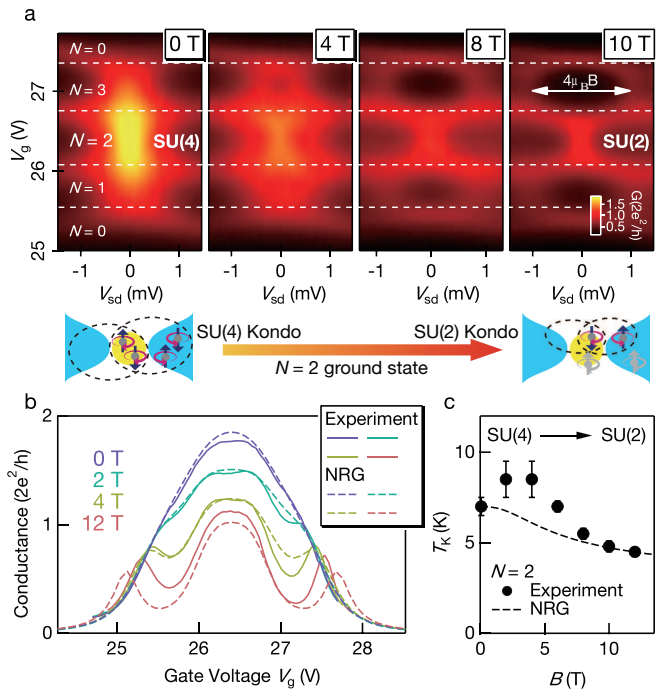}
\caption{(a)  Stability diagrams for $B=0$, 4, 8, and 10~T at $T=16$~mK as contour plots of the conductance $G$ as a function of $V_g$ and $V_{\rm sd}$. The Kondo resonance produces the bright broad vertical lines at $V_{\rm sd}=0$.  This ridge disappears at high field for $N=1$ and $3$. It is split into two satellite peaks at high $V_{\rm sd}$ separated by $\Delta eV_{\rm sd}\approx 4\mu_BB$. At $N=2$, $G$ decreases but it remains maximum at $V_{\rm sd}=0$. (b) Comparison of zero-bias conductance between the experiment (solid lines) and the NRG calculation (dashed lines) for several magnetic fields. (c) Field dependence of $T_K$ at $N=2$. The dashed line is the result of the NRG calculation multiplied by a factor of $1.3$ to fit the experimental value at $B=0$. The error bars result from the scaling procedure.}
\end{figure}

As seen in Fig.~2(a), when the field increases from 0 to 10~T, those ridges at $N=1$ and $N=3$ progressively disappear whereas the $N=2$ ridge remains until very high field. Indeed, on the contour plot, the bright vertical lines at $N=1$ and $N=3$ become dark, denoting that the resonance splits and $G$ becomes minimum at $V_{\rm sd}=0$. As indicated in the rightmost panel of Fig.~2(a), two satellite peaks appear at voltages close to $eV_{\rm sd}=\pm2\mu_BB$ as expected from the single-particle levels in Fig.~1(a). On the other hand, for $N=2$, the vertical Kondo ridge survives albeit with reduced intensity. At $B=0$, $G=1.85 G_Q$, which is close to $2G_Q$, the expected value for the $N=2$ unitary $SU(4)$ state. As $B$ increases, $G$ decreases toward $G_Q$, the value expected for the SU(2) Kondo effect~\cite{VanderWiel2000}.

{\it Comparison with NRG calculations.}---The above interpretation is quantitatively supported by the  numerical renormalization group (NRG)  calculations~\cite{IzumidaJPSJ1998,GalpinPRB2010,Teratani2016,SupMat}. We have successfully reproduced the complete shape of the zero-bias conductance $G$  as a function of $V_g$ as displayed in Fig.~2b. The solid lines are experimental results for $B=0$, 2, 4, and 12~T, whereas the dashed lines correspond to the NRG calculation for the same magnetic fields using the parameters $U/\Gamma=3.15$ ($U$ is the charging energy,  $\Gamma$ is the coupling strength) and $g_{\rm orb}\cos\theta=1$. This successful comparison allowed us to compute $T_K$ and the Wilson ratio $R$~\cite{Wilson1975}  from the NRG parameters as shown in Fig.~2(c) and Fig.~3(a), respectively. This ratio, proportional to the spin susceptibility, serves to quantify spin fluctuations on the dot, which create the Kondo effect (see Ref.~\cite{SupMat} for the definition of the present Wilson ratio). $R$ ranges from $1$ for noninteracting quasiparticles to $2$ for the $SU(2)$ Kondo state. Thus, Fig.~3(a) shows how quantum fluctuations continuously increase along the crossover when the magnetic field increases. The computed  $T_K$ is in good agreement with experimental values in the two limiting cases $SU(4)$ and $SU(2)$ as shown in Fig.~2(c)~\cite{SupMat}.  The decrease of $T_K$ in the $SU(2)$ state denotes an enhancement of the lifetime ($\frac{h}{k_BT_K}$) of the Kondo resonance due to the increase of the fluctuations.

\begin{figure}
\center
  \includegraphics[width=0.8\linewidth]{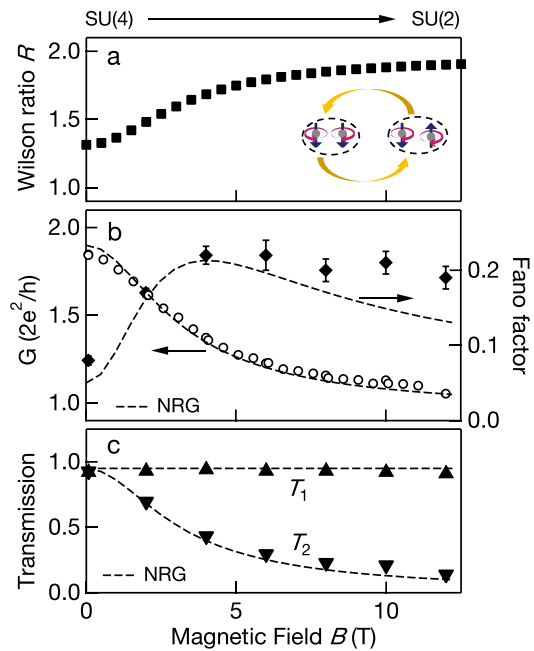}
\caption{ (a) Wilson ratio $R$ as a function of $B$. This ratio measures the strength of dominant quantum fluctuations between the two states represented on the graph, which yield the crossover~\cite{SupMat}. $R$ is computed by NRG calculations with parameters yielding the successful comparison with experiments for $G$ and $F$ at the filling $N=2$. 
(b) Conductance $G$ and Fano factor $F$ as a function of $B$. The $F$'s are obtained at each field by a linear fit of the current noise at very low current using the formula $S_I=2eFI_{\rm sd}$. The dashed lines are the corresponding NRG results.  (c) Magnetic field dependence of the two transport channels $T_1$ and $T_2$  (symbols), which  are calculated from $G$ and $F$ as explained in the text. The dashed lines are the corresponding NRG results.}
\end{figure}

{\it Shot noise along the symmetry crossover.}---In the rest of this Letter, we focus on the current noise $S_I$ to further clarify the symmetry crossover at $N=2$. 

First, we investigate the behavior of the transport channels along the crossover. For a Kondo dot, the linear transport properties are well described by noninteracting quasiparticles as demonstrated recently~\cite{Ferrier2015a}; $G$ and $S_I$  are expressed from the transmission channels $T_i$ via $G=G_Q\sum T_i$ and $S_I=2eF|I_{\rm sd}|$ where  $F=\sum_{i}T_i(1-T_i)/\sum_i T_i$ is the Fano factor ($i=1, 2,...$)~\cite{Blanter2000}. Assuming two independent channels, we obtain $T_{1,2}=g/2\pm1/2\sqrt{2g(1-F)-g^2}$, where $g=G/G_Q$. 

$G$ and $F$ measured exactly at $N=2$ are plotted as a function of $B$ in Fig.~3(b). The transmission probabilities $T_{1,2}$ deduced from $G$ and $F$ are plotted with symbols in Fig.~3(c). This figure shows that one transmission ($T_1$) remains almost unity during the crossover whereas the other ($T_2$) progressively vanishes to zero. This is the experimental evidence that the device undergoes a crossover from two perfect channels [$T_1 = T_2 \simeq 1$, hallmark of the $SU(4)$ Kondo state] to one perfect channel [$T_1\simeq 1$ and $T_2 \ll 1$, signature of the $SU(2)$ Kondo state]. The corresponding NRG results superposed by the dashed lines in Fig.~3c nicely reproduce these experimental findings. In this comparison a left-right asymmetry of the dot $G/2G_Q=0.95$ is taken into account for each transmission obtained from the NRG calculation. 

{\it Effective charge along the symmetry crossover.}---Now we present the central result, namely the direct measure of  quantum fluctuations along the crossover at $N=2$. At finite voltage, fluctuations induce a two particle scattering, which creates a backscattered current resulting from events involving one or two particles, or ``bubbles"~\cite{Sela2006}. Hence, the nonlinear noise is characterized by the effective charge $e^*$, the average backscattered charge, defined by~\cite{Sela2006}
\begin{equation}
e^*=\frac{e^2P_1+(2e)^2P_2}{eP_1+(2e)P_2},
\end{equation}
where $P_1$ is the probability of single-particle backscattering and $P_2$ that for the two-particle backscattering. At $T=0$, $e^*$ is experimentally defined as $S_K=2e^*|I_K|$. $S_K \equiv S_I-2eF|I_{\rm sd}|$ is the nonlinear part of the noise and $I_K \equiv G(0)V_{\rm sd}-I_{\rm sd}$ the nonlinear part of the current. Theoretically, $e^*/e=5/3$ in the $SU(2)$ state and $e^*/e=3/2$ in the $SU(4)$ state~\cite{Mora2009,Sakano2011a}. These values reflect that $P_2=P_1$ in the $SU(2)$ symmetry, whereas the two-particle backscattering  in the $SU(4)$ symmetry is less frequent such that $P_2=P_1/2$ because of the weakening of the fluctuations in the dot~\cite{Sakano2011a}. 

\begin{figure}
\center
  \includegraphics[width=1\linewidth]{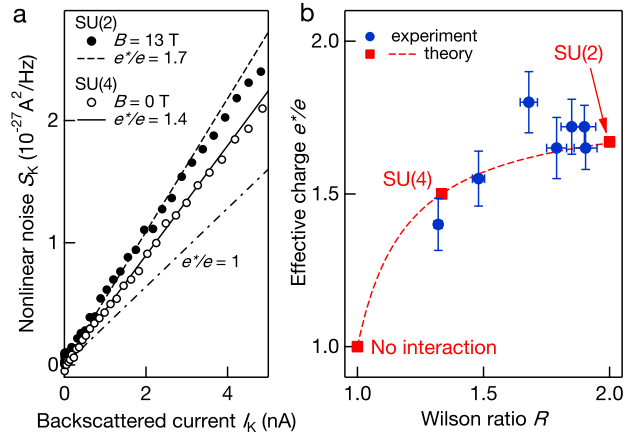}
\caption{ (a) Nonlinear noise $S_K$ as a function of the backscattered current $I_K$ at $B=0$~T [$SU(4)$ state] and $B=13$~T [$SU(2)$ state]. The solid and dashed  lines are the result of the linear fit yielding $e^*/e = 1.4 \pm 0.1$ at $B=0$~T and $e^*/e = 1.7 \pm 0.1$ at $B=13$~T, respectively. (b) The filled circles show the effective charge $e^*/e$ as a function of $R$, which quantifies the strength of fluctuations (the error bars for $R$ originate from the uncertainty of $g_{\rm orb}\cos\theta$). The three square symbols represent the theoretical prediction for $SU(4)$, $SU(2)$, and noninteracting particles. The dashed line is the  extended theoretical prediction given in the text [Eqn.~(\ref{effective_charge_wilson})].}
\end{figure}

Figure 4(a) represents $S_K$ as a function of $I_K$ measured at $B=0$~T [$SU(4)$ state] and at $B=13\ $T [$SU(2)$ state] at 16~mK. The lines are the linear fit which yield $e^*/e=1.4\pm 0.1$ for $SU(4)$ and $e^*/e=1.7\pm 0.1$ for $SU(2)$, being in good agreement with the theory. We found that $e^*$ almost continuously increases as $B$ increases from 0 to 13~T. This rules out $SU(2)\otimes SU(2)$ symmetry at $B=0$ instead of $SU(4)$, because $e^*$ would be independent of $B$ for this symmetry~\cite{SupMat}. Knowing $R$ as a function of $B$ [Fig.~3(a)], we represent $e^*$ as a function of $R$ in Fig.~4(b). Clearly, the effective charge gradually increases as $R$ increases. This graph illustrates how quantum fluctuations (namely, $R$) affect the two particle scattering ($e^*$). This is the key result of this Letter. It demonstrates that $e^*$ is a relevant experimental measure to quantify quantum fluctuations. 

For a well defined symmetry $SU(n)$, theory predicts that the Wilson ratio and the effective charge in the Kondo region are given only by $n$ such that $R =1+ \frac{1}{n-1}$ and $\frac{e^*}{e} = \frac{n+8}{n+4}$, respectively~\cite{Sakano2011}. We extend this relation in the broken symmetry region, yielding
\begin{equation}
\frac{e^*}{e}=\frac{1+9(R-1)}{1+5(R-1)},
\label{effective_charge_wilson}
\end{equation}
which is superposed as a dashed line in Fig.~4(b). Interestingly, our result is well reproduced by this relation continuously along the crossover from $SU(4)$ to $SU(2)$ even in the intermediate symmetry region. This emphasizes how nonequilibrium properties ($e^*$) and equilibrium quantities ($R$) are intricately linked in quantum many-body states.

{\it Conclusion.}---We addressed the microscopic mechanism along the crossover between Kondo many-body states with different symmetries by tuning the spin-orbital degeneracy in a CNT quantum dot  with a magnetic field at constant filling. The direct monitoring of the evolution of transmission channels by the conductance and the shot noise measurements demonstrates the continuous symmetry crossover between $SU(4)$ and $SU(2)$.
The effective charge and the Wilson ratio reveal that the enhancement of quantum fluctuations from SU(4) to SU(2) increases the lifetime of the Kondo resonance as well as the backscattering of pairs of entangled particles induced by the residual interaction. 
More generally, this work has experimentally established a convincing link between nonlinear noise and quantum fluctuations, which calls for an extension of the theory for ground states with broken symmetries. Hence, it paves the way towards the investigation of  the role of  fluctuations in quantum phase transitions, one of the main issues  in many-body physics. 

We thank H{\'e}l{\`e}ne Bouchiat and Sophie Gu{\'e}ron for a careful reading of the Letter. This work was partially supported by by Japan Society for the Promotion of Science KAKENHI Grant No.~JP26220711, No.~JP26400319, No.~JP16K17723,  No.~JP15K17680, No.~JP25103003, and No.~JP15H05854, the Yazaki Memorial Foundation for Science and Technology, Research Institute of Electrical Communication, Tohoku University, the French program Agence nationale de la Recherche DYMESYS (ANR2011-IS04-001-01), and Agence nationale de la Recherche MASH (ANR-12-BS04-0016).


\begin{thebibliography}{36}%
\makeatletter
\providecommand \@ifxundefined [1]{%
 \@ifx{#1\undefined}
}%
\providecommand \@ifnum [1]{%
 \ifnum #1\expandafter \@firstoftwo
 \else \expandafter \@secondoftwo
 \fi
}%
\providecommand \@ifx [1]{%
 \ifx #1\expandafter \@firstoftwo
 \else \expandafter \@secondoftwo
 \fi
}%
\providecommand \natexlab [1]{#1}%
\providecommand \enquote  [1]{``#1''}%
\providecommand \bibnamefont  [1]{#1}%
\providecommand \bibfnamefont [1]{#1}%
\providecommand \citenamefont [1]{#1}%
\providecommand \href@noop [0]{\@secondoftwo}%
\providecommand \href [0]{\begingroup \@sanitize@url \@href}%
\providecommand \@href[1]{\@@startlink{#1}\@@href}%
\providecommand \@@href[1]{\endgroup#1\@@endlink}%
\providecommand \@sanitize@url [0]{\catcode `\\12\catcode `\$12\catcode
  `\&12\catcode `\#12\catcode `\^12\catcode `\_12\catcode `\%12\relax}%
\providecommand \@@startlink[1]{}%
\providecommand \@@endlink[0]{}%
\providecommand \url  [0]{\begingroup\@sanitize@url \@url }%
\providecommand \@url [1]{\endgroup\@href {#1}{\urlprefix }}%
\providecommand \urlprefix  [0]{URL }%
\providecommand \Eprint [0]{\href }%
\providecommand \doibase [0]{http://dx.doi.org/}%
\providecommand \selectlanguage [0]{\@gobble}%
\providecommand \bibinfo  [0]{\@secondoftwo}%
\providecommand \bibfield  [0]{\@secondoftwo}%
\providecommand \translation [1]{[#1]}%
\providecommand \BibitemOpen [0]{}%
\providecommand \bibitemStop [0]{}%
\providecommand \bibitemNoStop [0]{.\EOS\space}%
\providecommand \EOS [0]{\spacefactor3000\relax}%
\providecommand \BibitemShut  [1]{\csname bibitem#1\endcsname}%
\let\auto@bib@innerbib\@empty
\bibitem [{\citenamefont {Kondo}(1964)}]{Kondo1964}%
  \BibitemOpen
  \bibfield  {author} {\bibinfo {author} {\bibfnamefont {J.}~\bibnamefont
  {Kondo}},\ }\href@noop {} {\bibfield  {journal} {\bibinfo  {journal} {Prog.
  Theor. Phys.}\ }\textbf {\bibinfo {volume} {32}},\ \bibinfo {pages} {37}
  (\bibinfo {year} {1964})}\BibitemShut {NoStop}%
\bibitem [{\citenamefont {Hewson}()}]{HewsonBook}%
  \BibitemOpen
  \bibfield  {author} {\bibinfo {author} {\bibfnamefont {A.~C.}\ \bibnamefont
  {Hewson}},\ }\href@noop {} {\emph {\bibinfo {title} {{The Kondo Problem to
  Heavy Fermions}}}},\ edited by\ \bibinfo {editor} {\bibfnamefont
  {D.}~\bibnamefont {Edwards}}\ and\ \bibinfo {editor} {\bibfnamefont
  {D.}~\bibnamefont {Melville}}\ (\bibinfo  {publisher} {Cambridge University
  Press})\BibitemShut {NoStop}%
\bibitem [{\citenamefont {Goldhaber-Gordon}\ \emph {et~al.}(1998)\citenamefont
  {Goldhaber-Gordon}, \citenamefont {Shtrikman}, \citenamefont {Mahalu},
  \citenamefont {Abusch-Magder}, \citenamefont {Meirav},\ and\ \citenamefont
  {Kastner}}]{Goldhaber-Gordon1998}%
  \BibitemOpen
  \bibfield  {author} {\bibinfo {author} {\bibfnamefont {D.}~\bibnamefont
  {Goldhaber-Gordon}}, \bibinfo {author} {\bibfnamefont {H.}~\bibnamefont
  {Shtrikman}}, \bibinfo {author} {\bibfnamefont {D.}~\bibnamefont {Mahalu}},
  \bibinfo {author} {\bibfnamefont {D.}~\bibnamefont {Abusch-Magder}}, \bibinfo
  {author} {\bibfnamefont {U.}~\bibnamefont {Meirav}}, \ and\ \bibinfo {author}
  {\bibfnamefont {M.~A.}\ \bibnamefont {Kastner}},\ }\href {\doibase
  10.1038/34373} {\bibfield  {journal} {\bibinfo  {journal} {Nature}\ }\textbf
  {\bibinfo {volume} {391}},\ \bibinfo {pages} {156} (\bibinfo {year}
  {1998})}\BibitemShut {NoStop}%
\bibitem [{\citenamefont {Cronenwett}\ \emph {et~al.}(1998)\citenamefont
  {Cronenwett}, \citenamefont {Oosterkamp},\ and\ \citenamefont
  {Kouwenhoven}}]{Cronenwett1998}%
  \BibitemOpen
  \bibfield  {author} {\bibinfo {author} {\bibfnamefont {S.~M.}\ \bibnamefont
  {Cronenwett}}, \bibinfo {author} {\bibfnamefont {T.~H.}\ \bibnamefont
  {Oosterkamp}}, \ and\ \bibinfo {author} {\bibfnamefont {L.~P.}\ \bibnamefont
  {Kouwenhoven}},\ }\href {\doibase 10.1126/science.281.5376.540} {\bibfield
  {journal} {\bibinfo  {journal} {Science}\ }\textbf {\bibinfo {volume}
  {281}},\ \bibinfo {pages} {540} (\bibinfo {year} {1998})}\BibitemShut
  {NoStop}%
\bibitem [{\citenamefont {Schmid}\ \emph {et~al.}(1998)\citenamefont {Schmid},
  \citenamefont {Weis}, \citenamefont {Eberl},\ and\ \citenamefont
  {v.~Klitzing}}]{Schmid1998}%
  \BibitemOpen
  \bibfield  {author} {\bibinfo {author} {\bibfnamefont {J.}~\bibnamefont
  {Schmid}}, \bibinfo {author} {\bibfnamefont {J.}~\bibnamefont {Weis}},
  \bibinfo {author} {\bibfnamefont {K.}~\bibnamefont {Eberl}}, \ and\ \bibinfo
  {author} {\bibfnamefont {K.}~\bibnamefont {v.~Klitzing}},\ }\href {\doibase
  10.1016/S0921-4526(98)00533-X} {\bibfield  {journal} {\bibinfo  {journal}
  {Phys. B Condens. Matter}\ }\textbf {\bibinfo {volume} {256-258}},\ \bibinfo
  {pages} {182} (\bibinfo {year} {1998})}\BibitemShut {NoStop}%
\bibitem [{\citenamefont {Coqblin}\ and\ \citenamefont
  {Schrieffer}(1969)}]{Coqblin1969}%
  \BibitemOpen
  \bibfield  {author} {\bibinfo {author} {\bibfnamefont {B.}~\bibnamefont
  {Coqblin}}\ and\ \bibinfo {author} {\bibfnamefont {J.~R.}\ \bibnamefont
  {Schrieffer}},\ }\href {\doibase 10.1103/PhysRev.185.847} {\bibfield
  {journal} {\bibinfo  {journal} {Phys. Rev.}\ }\textbf {\bibinfo {volume}
  {185}},\ \bibinfo {pages} {847} (\bibinfo {year} {1969})}\BibitemShut
  {NoStop}%
\bibitem [{\citenamefont {{Le Hur}}\ \emph {et~al.}(2007)\citenamefont {{Le
  Hur}}, \citenamefont {Simon},\ and\ \citenamefont {Loss}}]{LeHur2007}%
  \BibitemOpen
  \bibfield  {author} {\bibinfo {author} {\bibfnamefont {K.}~\bibnamefont {{Le
  Hur}}}, \bibinfo {author} {\bibfnamefont {P.}~\bibnamefont {Simon}}, \ and\
  \bibinfo {author} {\bibfnamefont {D.}~\bibnamefont {Loss}},\ }\href {\doibase
  10.1103/PhysRevB.75.035332} {\bibfield  {journal} {\bibinfo  {journal} {Phys.
  Rev. B}\ }\textbf {\bibinfo {volume} {75}},\ \bibinfo {pages} {035332}
  (\bibinfo {year} {2007})}\BibitemShut {NoStop}%
\bibitem [{\citenamefont {Nozi{\`{e}}res}\ and\ \citenamefont
  {Blandin}(1980)}]{Nozieres1980}%
  \BibitemOpen
  \bibfield  {author} {\bibinfo {author} {\bibfnamefont {P.}~\bibnamefont
  {Nozi{\`{e}}res}}\ and\ \bibinfo {author} {\bibfnamefont {A.}~\bibnamefont
  {Blandin}},\ }\href@noop {} {\bibfield  {journal} {\bibinfo  {journal} {J.
  Phys. (Paris)}\ }\textbf {\bibinfo {volume} {41}},\ \bibinfo {pages} {193}
  (\bibinfo {year} {1980})}\BibitemShut {NoStop}%
\bibitem [{\citenamefont {Nyg{\aa}rd}\ \emph {et~al.}(2000)\citenamefont
  {Nyg{\aa}rd}, \citenamefont {Cobden},\ and\ \citenamefont
  {Lindelof}}]{NygardNature2000}%
  \BibitemOpen
  \bibfield  {author} {\bibinfo {author} {\bibfnamefont {J.}~\bibnamefont
  {Nyg{\aa}rd}}, \bibinfo {author} {\bibfnamefont {D.~H.}\ \bibnamefont
  {Cobden}}, \ and\ \bibinfo {author} {\bibfnamefont {P.~E.}\ \bibnamefont
  {Lindelof}},\ }\href {http://dx.doi.org/10.1038/35042545} {\bibfield
  {journal} {\bibinfo  {journal} {Nature}\ }\textbf {\bibinfo {volume} {408}},\
  \bibinfo {pages} {342} (\bibinfo {year} {2000})}\BibitemShut {NoStop}%
\bibitem [{\citenamefont {Mora}\ \emph {et~al.}(2009)\citenamefont {Mora},
  \citenamefont {Vitushinsky}, \citenamefont {Leyronas}, \citenamefont
  {Clerk},\ and\ \citenamefont {{Le Hur}}}]{Mora2009}%
  \BibitemOpen
  \bibfield  {author} {\bibinfo {author} {\bibfnamefont {C.}~\bibnamefont
  {Mora}}, \bibinfo {author} {\bibfnamefont {P.}~\bibnamefont {Vitushinsky}},
  \bibinfo {author} {\bibfnamefont {X.}~\bibnamefont {Leyronas}}, \bibinfo
  {author} {\bibfnamefont {A.}~\bibnamefont {Clerk}}, \ and\ \bibinfo {author}
  {\bibfnamefont {K.}~\bibnamefont {{Le Hur}}},\ }\href {\doibase
  10.1103/PhysRevB.80.155322} {\bibfield  {journal} {\bibinfo  {journal} {Phys.
  Rev. B}\ }\textbf {\bibinfo {volume} {80}},\ \bibinfo {pages} {155322}
  (\bibinfo {year} {2009})}\BibitemShut {NoStop}%
\bibitem [{\citenamefont {Sakano}\ \emph
  {et~al.}(2011{\natexlab{a}})\citenamefont {Sakano}, \citenamefont {Fujii},\
  and\ \citenamefont {Oguri}}]{Sakano2011}%
  \BibitemOpen
  \bibfield  {author} {\bibinfo {author} {\bibfnamefont {R.}~\bibnamefont
  {Sakano}}, \bibinfo {author} {\bibfnamefont {T.}~\bibnamefont {Fujii}}, \
  and\ \bibinfo {author} {\bibfnamefont {A.}~\bibnamefont {Oguri}},\ }\href
  {\doibase 10.1103/PhysRevB.83.075440} {\bibfield  {journal} {\bibinfo
  {journal} {Phys. Rev. B}\ }\textbf {\bibinfo {volume} {83}},\ \bibinfo
  {pages} {075440} (\bibinfo {year} {2011}{\natexlab{a}})}\BibitemShut
  {NoStop}%
\bibitem [{\citenamefont {Sakano}\ and\ \citenamefont
  {Kawakami}(2006)}]{Sakano2006}%
  \BibitemOpen
  \bibfield  {author} {\bibinfo {author} {\bibfnamefont {R.}~\bibnamefont
  {Sakano}}\ and\ \bibinfo {author} {\bibfnamefont {N.}~\bibnamefont
  {Kawakami}},\ }\href {\doibase 10.1103/PhysRevB.73.155332} {\bibfield
  {journal} {\bibinfo  {journal} {Phys. Rev. B}\ }\textbf {\bibinfo {volume}
  {73}},\ \bibinfo {pages} {155332} (\bibinfo {year} {2006})}\BibitemShut
  {NoStop}%
\bibitem [{\citenamefont {Filippone}\ \emph {et~al.}(2014)\citenamefont
  {Filippone}, \citenamefont {Moca}, \citenamefont {Zar\'and},\ and\
  \citenamefont {Mora}}]{Filippone2014a}%
  \BibitemOpen
  \bibfield  {author} {\bibinfo {author} {\bibfnamefont {M.}~\bibnamefont
  {Filippone}}, \bibinfo {author} {\bibfnamefont {C.~P.}\ \bibnamefont {Moca}},
  \bibinfo {author} {\bibfnamefont {G.}~\bibnamefont {Zar\'and}}, \ and\
  \bibinfo {author} {\bibfnamefont {C.}~\bibnamefont {Mora}},\ }\href {\doibase
  10.1103/PhysRevB.90.121406} {\bibfield  {journal} {\bibinfo  {journal} {Phys.
  Rev. B}\ }\textbf {\bibinfo {volume} {90}},\ \bibinfo {pages} {121406}
  (\bibinfo {year} {2014})}\BibitemShut {NoStop}%
\bibitem [{\citenamefont {Gogolin}\ and\ \citenamefont
  {Komnik}(2006)}]{Gogolin2006}%
  \BibitemOpen
  \bibfield  {author} {\bibinfo {author} {\bibfnamefont {A.~O.}\ \bibnamefont
  {Gogolin}}\ and\ \bibinfo {author} {\bibfnamefont {A.}~\bibnamefont
  {Komnik}},\ }\href {\doibase 10.1103/PhysRevLett.97.016602} {\bibfield
  {journal} {\bibinfo  {journal} {Phys. Rev. Lett.}\ }\textbf {\bibinfo
  {volume} {97}},\ \bibinfo {pages} {016602} (\bibinfo {year}
  {2006})}\BibitemShut {NoStop}%
\bibitem [{\citenamefont {Sela}\ \emph {et~al.}(2006)\citenamefont {Sela},
  \citenamefont {Oreg}, \citenamefont {von Oppen},\ and\ \citenamefont
  {Koch}}]{Sela2006}%
  \BibitemOpen
  \bibfield  {author} {\bibinfo {author} {\bibfnamefont {E.}~\bibnamefont
  {Sela}}, \bibinfo {author} {\bibfnamefont {Y.}~\bibnamefont {Oreg}}, \bibinfo
  {author} {\bibfnamefont {F.}~\bibnamefont {von Oppen}}, \ and\ \bibinfo
  {author} {\bibfnamefont {J.}~\bibnamefont {Koch}},\ }\href {\doibase
  10.1103/PhysRevLett.97.086601} {\bibfield  {journal} {\bibinfo  {journal}
  {Phys. Rev. Lett.}\ }\textbf {\bibinfo {volume} {97}},\ \bibinfo {pages}
  {086601} (\bibinfo {year} {2006})}\BibitemShut {NoStop}%
\bibitem [{\citenamefont {Sasaki}\ \emph {et~al.}(2004)\citenamefont {Sasaki},
  \citenamefont {Amaha}, \citenamefont {Asakawa}, \citenamefont {Eto},\ and\
  \citenamefont {Tarucha}}]{SasakiPRL2004}%
  \BibitemOpen
  \bibfield  {author} {\bibinfo {author} {\bibfnamefont {S.}~\bibnamefont
  {Sasaki}}, \bibinfo {author} {\bibfnamefont {S.}~\bibnamefont {Amaha}},
  \bibinfo {author} {\bibfnamefont {N.}~\bibnamefont {Asakawa}}, \bibinfo
  {author} {\bibfnamefont {M.}~\bibnamefont {Eto}}, \ and\ \bibinfo {author}
  {\bibfnamefont {S.}~\bibnamefont {Tarucha}},\ }\href
  {http://link.aps.org/doi/10.1103/PhysRevLett.93.017205} {\bibfield  {journal}
  {\bibinfo  {journal} {Physical Review Letters}\ }\textbf {\bibinfo {volume}
  {93}},\ \bibinfo {pages} {017205} (\bibinfo {year} {2004})}\BibitemShut
  {NoStop}%
\bibitem [{\citenamefont {Choi}\ \emph {et~al.}(2005)\citenamefont {Choi},
  \citenamefont {L\'opez},\ and\ \citenamefont {Aguado}}]{ChoiPRL2005}%
  \BibitemOpen
  \bibfield  {author} {\bibinfo {author} {\bibfnamefont {M.-S.}\ \bibnamefont
  {Choi}}, \bibinfo {author} {\bibfnamefont {R.}~\bibnamefont {L\'opez}}, \
  and\ \bibinfo {author} {\bibfnamefont {R.}~\bibnamefont {Aguado}},\ }\href
  {\doibase 10.1103/PhysRevLett.95.067204} {\bibfield  {journal} {\bibinfo
  {journal} {Phys. Rev. Lett.}\ }\textbf {\bibinfo {volume} {95}},\ \bibinfo
  {pages} {067204} (\bibinfo {year} {2005})}\BibitemShut {NoStop}%
\bibitem [{\citenamefont {Jarillo-Herrero}\ \emph {et~al.}(2005)\citenamefont
  {Jarillo-Herrero}, \citenamefont {Kong}, \citenamefont {van~der Zant},
  \citenamefont {Dekker}, \citenamefont {Kouwenhoven},\ and\ \citenamefont
  {De~Franceschi}}]{PabloNature2005}%
  \BibitemOpen
  \bibfield  {author} {\bibinfo {author} {\bibfnamefont {P.}~\bibnamefont
  {Jarillo-Herrero}}, \bibinfo {author} {\bibfnamefont {J.}~\bibnamefont
  {Kong}}, \bibinfo {author} {\bibfnamefont {H.~S.~J.}\ \bibnamefont {van~der
  Zant}}, \bibinfo {author} {\bibfnamefont {C.}~\bibnamefont {Dekker}},
  \bibinfo {author} {\bibfnamefont {L.~P.}\ \bibnamefont {Kouwenhoven}}, \ and\
  \bibinfo {author} {\bibfnamefont {S.}~\bibnamefont {De~Franceschi}},\ }\href
  {\doibase
  http://www.nature.com/nature/journal/v434/n7032/suppinfo/nature03422_S1.html}
  {\bibfield  {journal} {\bibinfo  {journal} {Nature}\ }\textbf {\bibinfo
  {volume} {434}},\ \bibinfo {pages} {484} (\bibinfo {year}
  {2005})}\BibitemShut {NoStop}%
\bibitem [{\citenamefont {Delattre}\ \emph {et~al.}(2009)\citenamefont
  {Delattre}, \citenamefont {Feuillet-Palma}, \citenamefont {Herrmann},
  \citenamefont {Morfin}, \citenamefont {Berroir}, \citenamefont {F{\`e}ve},
  \citenamefont {Pla{\c c}ais}, \citenamefont {Glattli}, \citenamefont {Choi},
  \citenamefont {Mora},\ and\ \citenamefont {Kontos}}]{DelattreNatPhys2009}%
  \BibitemOpen
  \bibfield  {author} {\bibinfo {author} {\bibfnamefont {T.}~\bibnamefont
  {Delattre}}, \bibinfo {author} {\bibfnamefont {C.}~\bibnamefont
  {Feuillet-Palma}}, \bibinfo {author} {\bibfnamefont {L.~G.}\ \bibnamefont
  {Herrmann}}, \bibinfo {author} {\bibfnamefont {P.}~\bibnamefont {Morfin}},
  \bibinfo {author} {\bibfnamefont {J.~M.}\ \bibnamefont {Berroir}}, \bibinfo
  {author} {\bibfnamefont {G.}~\bibnamefont {F{\`e}ve}}, \bibinfo {author}
  {\bibfnamefont {B.}~\bibnamefont {Pla{\c c}ais}}, \bibinfo {author}
  {\bibfnamefont {D.~C.}\ \bibnamefont {Glattli}}, \bibinfo {author}
  {\bibfnamefont {M.~S.}\ \bibnamefont {Choi}}, \bibinfo {author}
  {\bibfnamefont {C.}~\bibnamefont {Mora}}, \ and\ \bibinfo {author}
  {\bibfnamefont {T.}~\bibnamefont {Kontos}},\ }\href {\doibase
  10.1038/nphys1186} {\bibfield  {journal} {\bibinfo  {journal} {Nature
  Physics}\ }\textbf {\bibinfo {volume} {5}},\ \bibinfo {pages} {208} (\bibinfo
  {year} {2009})}\BibitemShut {NoStop}%
\bibitem [{\citenamefont {Laird}\ \emph {et~al.}(2015)\citenamefont {Laird},
  \citenamefont {Kuemmeth}, \citenamefont {Steele}, \citenamefont
  {Grove-Rasmussen}, \citenamefont {Rd}, \citenamefont {Flensberg},\ and\
  \citenamefont {Kouwenhoven}}]{Laird2014a}%
  \BibitemOpen
  \bibfield  {author} {\bibinfo {author} {\bibfnamefont {E.~A.}\ \bibnamefont
  {Laird}}, \bibinfo {author} {\bibfnamefont {F.}~\bibnamefont {Kuemmeth}},
  \bibinfo {author} {\bibfnamefont {G.}~\bibnamefont {Steele}}, \bibinfo
  {author} {\bibfnamefont {K.}~\bibnamefont {Grove-Rasmussen}}, \bibinfo
  {author} {\bibfnamefont {J.~N.}\ \bibnamefont {Rd}}, \bibinfo {author}
  {\bibfnamefont {K.}~\bibnamefont {Flensberg}}, \ and\ \bibinfo {author}
  {\bibfnamefont {L.~P.}\ \bibnamefont {Kouwenhoven}},\ }\href {\doibase
  10.1103/RevModPhys.87.703} {\bibfield  {journal} {\bibinfo  {journal} {Rev.
  Mod. Phys.}\ }\textbf {\bibinfo {volume} {87}},\ \bibinfo {pages} {703}
  (\bibinfo {year} {2015})}\BibitemShut {NoStop}%
\bibitem [{\citenamefont {Sakano}\ \emph
  {et~al.}(2011{\natexlab{b}})\citenamefont {Sakano}, \citenamefont {Oguri},
  \citenamefont {Kato},\ and\ \citenamefont {Tarucha}}]{Sakano2011a}%
  \BibitemOpen
  \bibfield  {author} {\bibinfo {author} {\bibfnamefont {R.}~\bibnamefont
  {Sakano}}, \bibinfo {author} {\bibfnamefont {A.}~\bibnamefont {Oguri}},
  \bibinfo {author} {\bibfnamefont {T.}~\bibnamefont {Kato}}, \ and\ \bibinfo
  {author} {\bibfnamefont {S.}~\bibnamefont {Tarucha}},\ }\href {\doibase
  10.1103/PhysRevB.83.241301} {\bibfield  {journal} {\bibinfo  {journal} {Phys.
  Rev. B}\ }\textbf {\bibinfo {volume} {83}},\ \bibinfo {pages} {241301}
  (\bibinfo {year} {2011}{\natexlab{b}})}\BibitemShut {NoStop}%
\bibitem [{\citenamefont {Oguri}(2001)}]{Oguri2001}%
  \BibitemOpen
  \bibfield  {author} {\bibinfo {author} {\bibfnamefont {A.}~\bibnamefont
  {Oguri}},\ }\href {\doibase 10.1103/PhysRevB.64.153305} {\bibfield  {journal}
  {\bibinfo  {journal} {Phys. Rev. B}\ }\textbf {\bibinfo {volume} {64}},\
  \bibinfo {pages} {153305} (\bibinfo {year} {2001})}\BibitemShut {NoStop}%
\bibitem [{\citenamefont {{Glazman}}\ and\ \citenamefont
  {{Pustilnik}}(2005)}]{Glazman2005condmat}%
  \BibitemOpen
  \bibfield  {author} {\bibinfo {author} {\bibfnamefont {L.~I.}\ \bibnamefont
  {{Glazman}}}\ and\ \bibinfo {author} {\bibfnamefont {M.}~\bibnamefont
  {{Pustilnik}}},\ }\href@noop {} {\  (\bibinfo {year} {2005})},\ \Eprint
  {http://arxiv.org/abs/cond-mat/0501007} {cond-mat/0501007} \BibitemShut
  {NoStop}%
\bibitem [{\citenamefont {Oguri}(2005)}]{Oguri2005a}%
  \BibitemOpen
  \bibfield  {author} {\bibinfo {author} {\bibfnamefont {A.}~\bibnamefont
  {Oguri}},\ }\href {\doibase 10.1143/JPSJ.74.110} {\bibfield  {journal}
  {\bibinfo  {journal} {J. Phys. Soc. Japan}\ }\textbf {\bibinfo {volume}
  {74}},\ \bibinfo {pages} {110} (\bibinfo {year} {2005})}\BibitemShut
  {NoStop}%
\bibitem [{\citenamefont {Momma}\ and\ \citenamefont
  {Izumi}(2011)}]{MommaJAC2011}%
  \BibitemOpen
  \bibfield  {author} {\bibinfo {author} {\bibfnamefont {K.}~\bibnamefont
  {Momma}}\ and\ \bibinfo {author} {\bibfnamefont {F.}~\bibnamefont {Izumi}},\
  }\href {\doibase 10.1107/S0021889811038970} {\bibfield  {journal} {\bibinfo
  {journal} {Journal of Applied Crystallography}\ }\textbf {\bibinfo {volume}
  {44}},\ \bibinfo {pages} {1272} (\bibinfo {year} {2011})}\BibitemShut
  {NoStop}%
\bibitem [{Sup()}]{SupMat}%
  \BibitemOpen
  \href@noop {} {\bibinfo  {journal} {See Supplemental Material 
at \url{http://link.aps.org/supplemental/10.1103/PhysRevLett.118.196803}, which
  includes details on the experimental and theoretical aspects of this work,
  such as the model for the NRG calculations, determination of the angle of the
  field, Wilson ratio, and so on.}\ }\BibitemShut {NoStop}%
\bibitem [{\citenamefont {Ferrier}\ \emph {et~al.}(2016)\citenamefont
  {Ferrier}, \citenamefont {Arakawa}, \citenamefont {Hata}, \citenamefont
  {Fujiwara}, \citenamefont {Delagrange}, \citenamefont {Weil}, \citenamefont
  {Deblock}, \citenamefont {Sakano}, \citenamefont {Oguri},\ and\ \citenamefont
  {Kobayashi}}]{Ferrier2015a}%
  \BibitemOpen
\bibfield  {journal} {  }\bibfield  {author} {\bibinfo {author} {\bibfnamefont
  {M.}~\bibnamefont {Ferrier}}, \bibinfo {author} {\bibfnamefont
  {T.}~\bibnamefont {Arakawa}}, \bibinfo {author} {\bibfnamefont
  {T.}~\bibnamefont {Hata}}, \bibinfo {author} {\bibfnamefont {R.}~\bibnamefont
  {Fujiwara}}, \bibinfo {author} {\bibfnamefont {R.}~\bibnamefont
  {Delagrange}}, \bibinfo {author} {\bibfnamefont {R.}~\bibnamefont {Weil}},
  \bibinfo {author} {\bibfnamefont {R.}~\bibnamefont {Deblock}}, \bibinfo
  {author} {\bibfnamefont {R.}~\bibnamefont {Sakano}}, \bibinfo {author}
  {\bibfnamefont {A.}~\bibnamefont {Oguri}}, \ and\ \bibinfo {author}
  {\bibfnamefont {K.}~\bibnamefont {Kobayashi}},\ }\href {\doibase
  10.1038/nphys3556} {\bibfield  {journal} {\bibinfo  {journal} {Nat. Phys.}\
  }\textbf {\bibinfo {volume} {12}},\ \bibinfo {pages} {230} (\bibinfo {year}
  {2016})}\BibitemShut {NoStop}%
\bibitem [{\citenamefont {Kasumov}\ \emph {et~al.}(2007)\citenamefont
  {Kasumov}, \citenamefont {Shailos}, \citenamefont {Khodos}, \citenamefont
  {Volkov}, \citenamefont {Levashov}, \citenamefont {Matveev}, \citenamefont
  {Gu{\'{e}}ron}, \citenamefont {Kobylko}, \citenamefont {Kociak},
  \citenamefont {Bouchiat}, \citenamefont {Agache}, \citenamefont {Rollier},
  \citenamefont {Buchaillot}, \citenamefont {Bonnot},\ and\ \citenamefont
  {Kasumov}}]{Kasumov2007}%
  \BibitemOpen
  \bibfield  {author} {\bibinfo {author} {\bibfnamefont {Y.}~\bibnamefont
  {Kasumov}}, \bibinfo {author} {\bibfnamefont {A.}~\bibnamefont {Shailos}},
  \bibinfo {author} {\bibfnamefont {I.}~\bibnamefont {Khodos}}, \bibinfo
  {author} {\bibfnamefont {V.}~\bibnamefont {Volkov}}, \bibinfo {author}
  {\bibfnamefont {V.}~\bibnamefont {Levashov}}, \bibinfo {author}
  {\bibfnamefont {V.}~\bibnamefont {Matveev}}, \bibinfo {author} {\bibfnamefont
  {S.}~\bibnamefont {Gu{\'{e}}ron}}, \bibinfo {author} {\bibfnamefont
  {M.}~\bibnamefont {Kobylko}}, \bibinfo {author} {\bibfnamefont
  {M.}~\bibnamefont {Kociak}}, \bibinfo {author} {\bibfnamefont
  {H.}~\bibnamefont {Bouchiat}}, \bibinfo {author} {\bibfnamefont
  {V.}~\bibnamefont {Agache}}, \bibinfo {author} {\bibfnamefont
  {A.}~\bibnamefont {Rollier}}, \bibinfo {author} {\bibfnamefont
  {L.}~\bibnamefont {Buchaillot}}, \bibinfo {author} {\bibfnamefont
  {A.}~\bibnamefont {Bonnot}}, \ and\ \bibinfo {author} {\bibfnamefont
  {A.}~\bibnamefont {Kasumov}},\ }\href {\doibase 10.1007/s00339-007-4028-3}
  {\bibfield  {journal} {\bibinfo  {journal} {Appl. Phys. A}\ }\textbf
  {\bibinfo {volume} {88}},\ \bibinfo {pages} {687} (\bibinfo {year}
  {2007})}\BibitemShut {NoStop}%
\bibitem [{\citenamefont {Arakawa}\ \emph {et~al.}(2013)\citenamefont
  {Arakawa}, \citenamefont {Nishihara}, \citenamefont {Maeda}, \citenamefont
  {Norimoto},\ and\ \citenamefont {Kobayashi}}]{Arakawa2013}%
  \BibitemOpen
  \bibfield  {author} {\bibinfo {author} {\bibfnamefont {T.}~\bibnamefont
  {Arakawa}}, \bibinfo {author} {\bibfnamefont {Y.}~\bibnamefont {Nishihara}},
  \bibinfo {author} {\bibfnamefont {M.}~\bibnamefont {Maeda}}, \bibinfo
  {author} {\bibfnamefont {S.}~\bibnamefont {Norimoto}}, \ and\ \bibinfo
  {author} {\bibfnamefont {K.}~\bibnamefont {Kobayashi}},\ }\href {\doibase
  10.1063/1.4826681} {\bibfield  {journal} {\bibinfo  {journal} {Appl. Phys.
  Lett.}\ }\textbf {\bibinfo {volume} {103}},\ \bibinfo {pages} {172104}
  (\bibinfo {year} {2013})}\BibitemShut {NoStop}%
\bibitem [{\citenamefont {Makarovski}\ \emph {et~al.}(2007)\citenamefont
  {Makarovski}, \citenamefont {Zhukov}, \citenamefont {Liu},\ and\
  \citenamefont {Finkelstein}}]{Makarovski2007}%
  \BibitemOpen
  \bibfield  {author} {\bibinfo {author} {\bibfnamefont {A.}~\bibnamefont
  {Makarovski}}, \bibinfo {author} {\bibfnamefont {A.}~\bibnamefont {Zhukov}},
  \bibinfo {author} {\bibfnamefont {J.}~\bibnamefont {Liu}}, \ and\ \bibinfo
  {author} {\bibfnamefont {G.}~\bibnamefont {Finkelstein}},\ }\href {\doibase
  10.1103/PhysRevB.75.241407} {\bibfield  {journal} {\bibinfo  {journal} {Phys.
  Rev. B}\ }\textbf {\bibinfo {volume} {75}},\ \bibinfo {pages} {241407}
  (\bibinfo {year} {2007})}\BibitemShut {NoStop}%
\bibitem [{\citenamefont {van~der Wiel}\ \emph {et~al.}(2000)\citenamefont
  {van~der Wiel}, \citenamefont {{De Franceschi}}, \citenamefont {Fujisawa},
  \citenamefont {Elzerman}, \citenamefont {Tarucha},\ and\ \citenamefont
  {Kouwenhoven}}]{VanderWiel2000}%
  \BibitemOpen
  \bibfield  {author} {\bibinfo {author} {\bibfnamefont {W.~G.}\ \bibnamefont
  {van~der Wiel}}, \bibinfo {author} {\bibfnamefont {S.}~\bibnamefont {{De
  Franceschi}}}, \bibinfo {author} {\bibfnamefont {T.}~\bibnamefont
  {Fujisawa}}, \bibinfo {author} {\bibfnamefont {J.}~\bibnamefont {Elzerman}},
  \bibinfo {author} {\bibfnamefont {S.}~\bibnamefont {Tarucha}}, \ and\
  \bibinfo {author} {\bibfnamefont {L.~P.}\ \bibnamefont {Kouwenhoven}},\
  }\href {\doibase 10.1126/science.289.5487.2105} {\bibfield  {journal}
  {\bibinfo  {journal} {Science}\ }\textbf {\bibinfo {volume} {289}},\ \bibinfo
  {pages} {2105} (\bibinfo {year} {2000})}\BibitemShut {NoStop}%
\bibitem [{\citenamefont {Izumida}\ \emph {et~al.}(1998)\citenamefont
  {Izumida}, \citenamefont {Sakai},\ and\ \citenamefont
  {Shimizu}}]{IzumidaJPSJ1998}%
  \BibitemOpen
  \bibfield  {author} {\bibinfo {author} {\bibfnamefont {W.}~\bibnamefont
  {Izumida}}, \bibinfo {author} {\bibfnamefont {O.}~\bibnamefont {Sakai}}, \
  and\ \bibinfo {author} {\bibfnamefont {Y.}~\bibnamefont {Shimizu}},\ }\href
  {\doibase 10.1143/JPSJ.67.2444} {\bibfield  {journal} {\bibinfo  {journal}
  {Journal of the Physical Society of Japan}\ }\textbf {\bibinfo {volume}
  {67}},\ \bibinfo {pages} {2444} (\bibinfo {year} {1998})}\BibitemShut
  {NoStop}%
\bibitem [{\citenamefont {Galpin}\ \emph {et~al.}(2010)\citenamefont {Galpin},
  \citenamefont {Jayatilaka}, \citenamefont {Logan},\ and\ \citenamefont
  {Anders}}]{GalpinPRB2010}%
  \BibitemOpen
  \bibfield  {author} {\bibinfo {author} {\bibfnamefont {M.~R.}\ \bibnamefont
  {Galpin}}, \bibinfo {author} {\bibfnamefont {F.~W.}\ \bibnamefont
  {Jayatilaka}}, \bibinfo {author} {\bibfnamefont {D.~E.}\ \bibnamefont
  {Logan}}, \ and\ \bibinfo {author} {\bibfnamefont {F.~B.}\ \bibnamefont
  {Anders}},\ }\href {\doibase 10.1103/PhysRevB.81.075437} {\bibfield
  {journal} {\bibinfo  {journal} {Phys. Rev. B}\ }\textbf {\bibinfo {volume}
  {81}},\ \bibinfo {pages} {075437} (\bibinfo {year} {2010})}\BibitemShut
  {NoStop}%
\bibitem [{\citenamefont {Teratani}\ \emph {et~al.}(2016)\citenamefont
  {Teratani}, \citenamefont {Sakano}, \citenamefont {Fujiwara}, \citenamefont
  {Hata}, \citenamefont {Arakawa}, \citenamefont {Ferrier}, \citenamefont
  {Kobayashi},\ and\ \citenamefont {Oguri}}]{Teratani2016}%
  \BibitemOpen
  \bibfield  {author} {\bibinfo {author} {\bibfnamefont {Y.}~\bibnamefont
  {Teratani}}, \bibinfo {author} {\bibfnamefont {R.}~\bibnamefont {Sakano}},
  \bibinfo {author} {\bibfnamefont {R.}~\bibnamefont {Fujiwara}}, \bibinfo
  {author} {\bibfnamefont {T.}~\bibnamefont {Hata}}, \bibinfo {author}
  {\bibfnamefont {T.}~\bibnamefont {Arakawa}}, \bibinfo {author} {\bibfnamefont
  {M.}~\bibnamefont {Ferrier}}, \bibinfo {author} {\bibfnamefont
  {K.}~\bibnamefont {Kobayashi}}, \ and\ \bibinfo {author} {\bibfnamefont
  {A.}~\bibnamefont {Oguri}},\ }\href@noop {} {\bibfield  {journal} {\bibinfo
  {journal} {J. Phys. Soc. Jpn.}\ }\textbf {\bibinfo {volume} {85}},\ \bibinfo
  {pages} {094718} (\bibinfo {year} {2016})}\BibitemShut {NoStop}%
\bibitem [{\citenamefont {Wilson}(1975)}]{Wilson1975}%
  \BibitemOpen
  \bibfield  {author} {\bibinfo {author} {\bibfnamefont {K.~G.}\ \bibnamefont
  {Wilson}},\ }\href {\doibase 10.1103/RevModPhys.47.773} {\bibfield  {journal}
  {\bibinfo  {journal} {Rev. Mod. Phys.}\ }\textbf {\bibinfo {volume} {47}},\
  \bibinfo {pages} {773} (\bibinfo {year} {1975})}\BibitemShut {NoStop}%
\bibitem [{\citenamefont {Blanter}\ and\ \citenamefont
  {B{\"u}ttiker}(2000)}]{Blanter2000}%
  \BibitemOpen
  \bibfield  {author} {\bibinfo {author} {\bibfnamefont {Y.~M.}\ \bibnamefont
  {Blanter}}\ and\ \bibinfo {author} {\bibfnamefont {M.}~\bibnamefont
  {B{\"u}ttiker}},\ }\href@noop {} {\bibfield  {journal} {\bibinfo  {journal}
  {Phys. Rep.}\ }\textbf {\bibinfo {volume} {336}},\ \bibinfo {pages} {1}
  (\bibinfo {year} {2000})}\BibitemShut {NoStop}%
\end{thebibliography}
\end{document}